\documentclass{moriond}

\usepackage{subcaption}
\usepackage{units}
\usepackage{mathrsfs}
\usepackage{mathtools}
\usepackage{bm}

\def\Journal#1#2#3#4{{#1} {\bf #2}, #3 (#4)}

\def\AA{\em A\&A}
\def\ADASS{\em Proceedings of XXXI Astronomical Data Analysis Software and Systems (ADASS) conference}
\def\APJ{\em Astrophys. J.}
\def\JCAP{\em JCAP}
\def\ICRCA{\em Proceedings of $36^{\text{th}} $ International Cosmic Ray Conference (ICRC)}
\def\ICRCB{\em Proceedings of $37^{\text{th}} $ International Cosmic Ray Conference (ICRC)}

\def\PDU{\em Physics of the Dark Universe}
\def\PLB{{\em Phys. Lett.}  B}
\def\PRL{\em Phys. Rev. Lett.}
\def\PRD{{\em Phys. Rev.} D}
\def\ZENODO{\em Zenodo}

\def\be{\begin{equation}}
\def\ee{\end{equation}}
\def\bea{\begin{eqnarray}}
\def\eea{\end{eqnarray}}

\newcommand{\Photo}{\includegraphics[height=35mm]{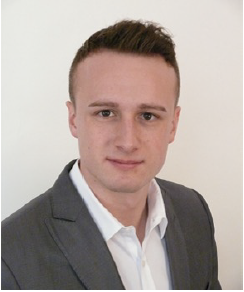}}
\newcommand{\email}[1]{%
  \href{mailto:#1}{\footnotesize\nolinkurl{#1}}
}%

\begin{document}
\vspace*{4cm}
\title{Constraining branon dark matter from observations of the dwarf spheroidal galaxies with the MAGIC telescopes}

\author{T. Miener$^{1}$, D. Nieto$^{1}$, V. Gammaldi$^{2}$, D.
Kerszberg$^{3}$, and J. Rico$^{3}$}

\address{$^{1}$Instituto de F\'{i}sica de Part\'{i}culas y del Cosmos and Department of EMFTEL, Universidad Complutense de Madrid, E-28040 Madrid, Spain; \email{tmiener@ucm.es}\\
$^{2}$Departamento de F\'{i}sica Te\'{o}rica, Universidad Aut\'{o}noma de Madrid, Madrid, Spain $\&$  Instituto de F\'{i}sica Te\'{o}rica, UAM/CSIC, E-28049 Madrid, Spain\\
$^{3}$Institut de F\'isica d'Altes Energies (IFAE), The Barcelona Institute of Science and Technology (BIST), E-08193 Bellaterra (Barcelona), Spain}

\maketitle\abstracts{
We present the first branon dark matter (DM) search in the very high-energy gamma-ray band with observations of the dwarf spheroidal galaxy Segue~1 carried out by the Major Atmospheric Gamma Imaging Cherenkov (MAGIC) telescope system. Branons are new degrees of freedom that appear in flexible brane-world models corresponding to brane fluctuations. They behave as weakly interacting massive particles, which are natural DM candidates. In the absence of a gamma-ray signal in the Segue~1 data, we place constraints on the branon DM parameter space by using a binned likelihood analysis. Our most constraining limit to the thermally-averaged annihilation cross-section (at $95\%$ confidence level) corresponds to $ \langle \sigma v \rangle \simeq \unit[1.4 \times 10^{-23}]{cm^{3}s^{-1}} $ at a branon DM mass of $ \sim \unit[0.7]{TeV}$.}

\section{Introduction}
Astrophysical and cosmological evidences propose that non-baryonic cold  dark matter (DM) accounts for $84\%$ of the matter density of the Universe~\cite{Aghanim:2018eyx}. Nevertheless, the nature of DM is still an open question for modern physics, as this elusive kind of matter can not be made of any of the Standard Model (SM) particles. Brane-world theory has been put forward as a prospective framework for DM candidates~\cite{2003PhRvL..90x1301C}. The characteristics of the suggested massive brane fluctuations (branons) in this theory match the ones of weakly interacting massive particles (WIMPs), which are one of the most favored candidates of cold DM~\cite{2012PhRvD..86b3506S}.

Dwarf spheroidal galaxies (dSphs) are preferred targets for indirect DM searches, due to their close proximities and high mass-to-light ratios. In this work, we are focusing on the ultra-faint dSph Segue~1 located in the Northern Hemisphere and outside of the Galactic plane ($ \text{RA} = \unit[10.12]{h} $, $ \text{DEC} = 16.08^{\circ}$) only $ \unit[23\pm2]{kpc}$ away from us~\cite{2007ApJ...654..897B}. The Segue~1 observational campaign was carried out by the ground-based gamma-ray telescope MAGIC between 2011 and 2013 and is with 157.9 hours the deepest observation of any dSph by a Cherenkov telescope to date~\cite{2014JCAP...02..008A}.

\section{Expected branon dark matter gamma-ray flux}

\begin{figure}[h]
    \centering
    \begin{subfigure}{.49\textwidth}
        \includegraphics[width=\textwidth]{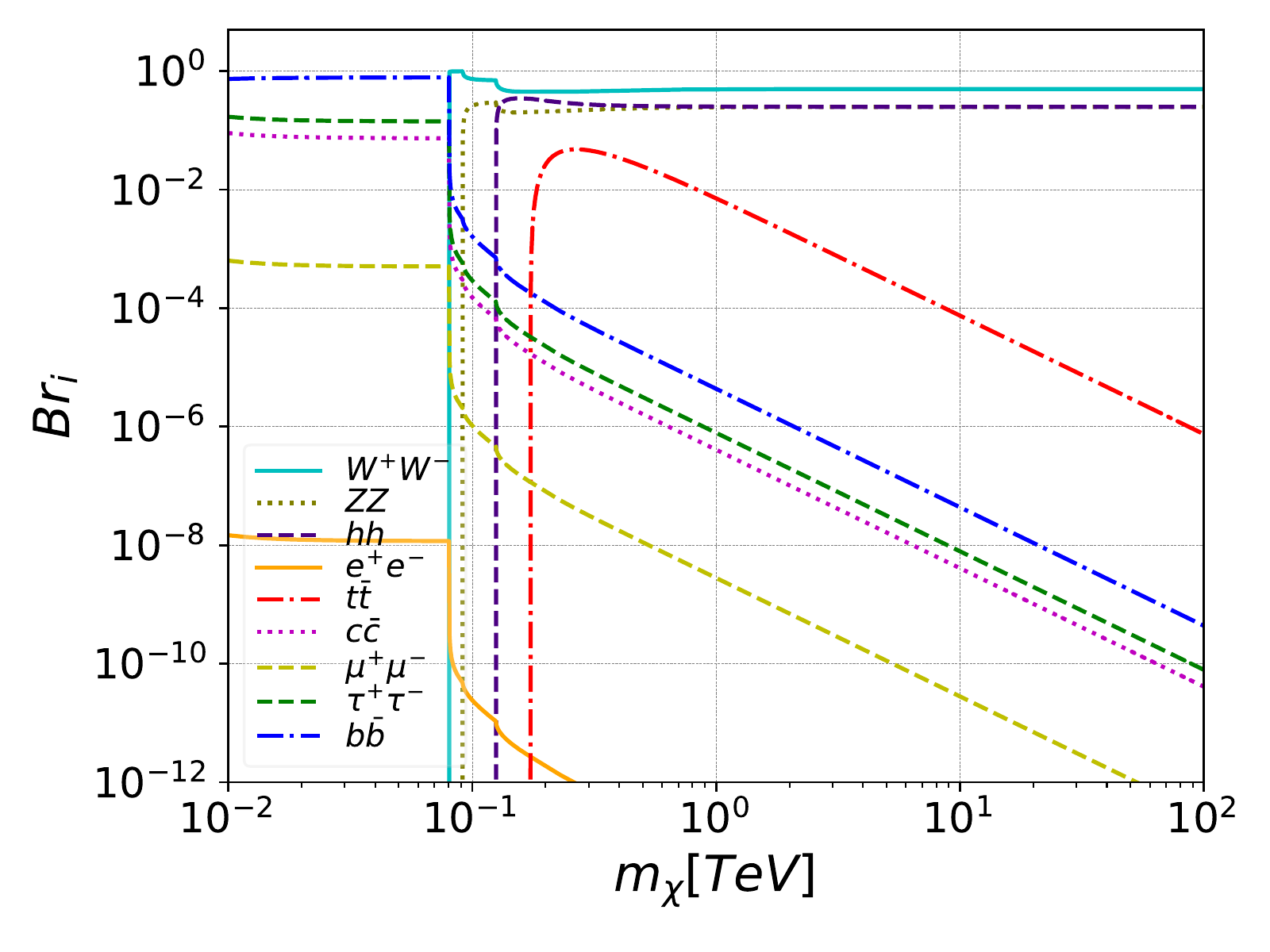}
    \end{subfigure}
    \begin{subfigure}{.49\textwidth}
        \includegraphics[width=\textwidth]{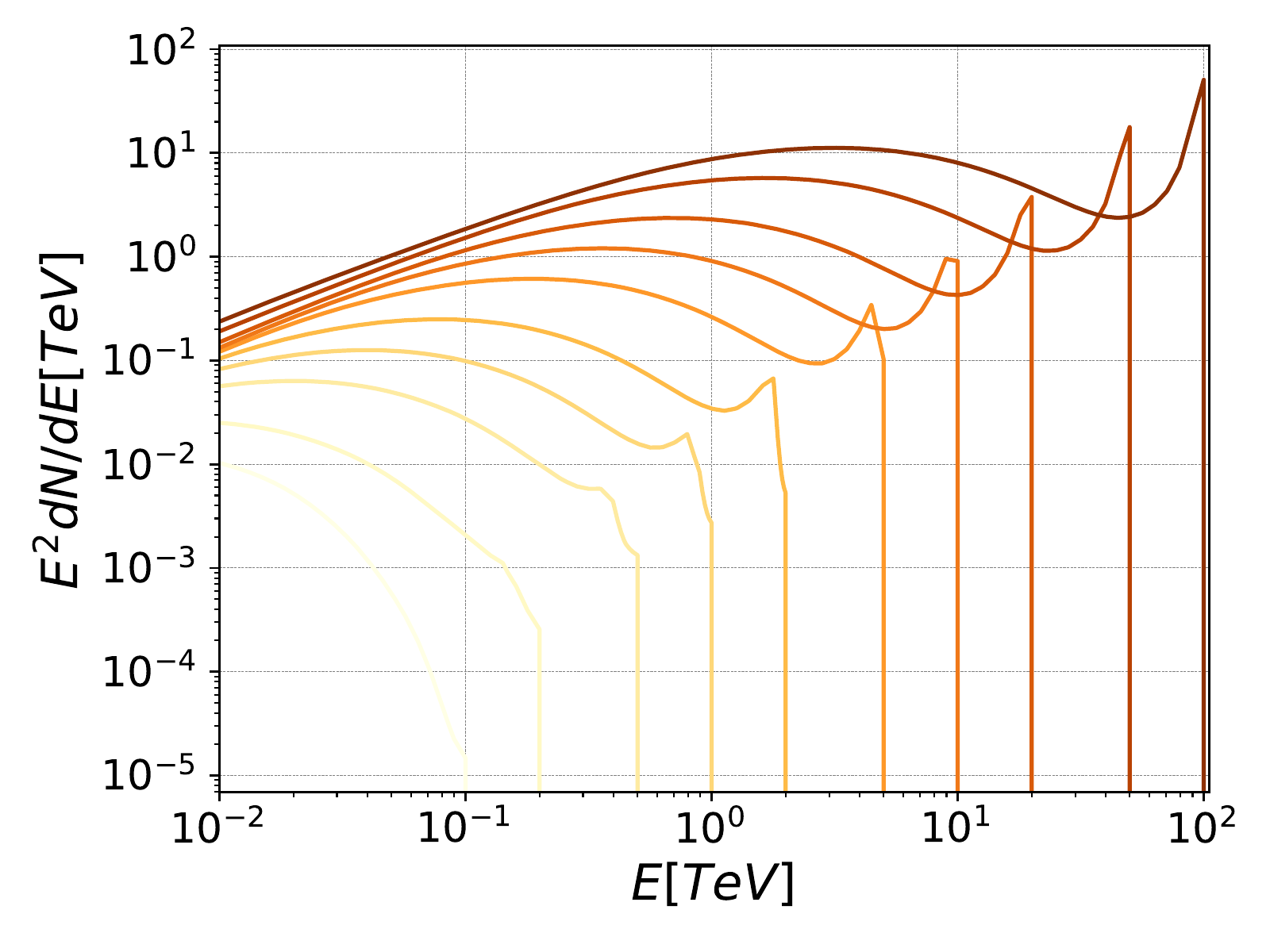}
    \end{subfigure}
\caption{Taken from JCAP05(2022)005. See text for more details.}
\label{fig:Branching_ratios_and_spectra}
\end{figure}

The expected gamma-ray flux produced by branon DM annihilation is composed of the \textit{astrophysical} factor (\textit{J}-factor), which depends on both the distance $ l $ and the DM distribution at the source region $ \rho_{\text{DM}} $, and the branon DM annihilation photon yield. It can be expressed in a given region of the sky, $\Delta\Omega$, and observed at Earth as
\begin{equation}
    \label{eq:Branon_Flux}
    \frac{\text{d}\Phi}{\text{d}E}\left( \langle\sigma v\rangle \right) = \underbrace{\int_{\Delta\Omega} d\Omega' \int_{\text{l.o.s.}} dl \, \rho_{\text{DM}}^{2} (l,\Omega')}_{\text{Astrophysics}} \cdot \underbrace{\frac{1}{4\pi} \frac{\langle\sigma v\rangle}{2m^{2}_{\chi}} \sum_{i=1}^{n} \text{Br}_{i} \frac{\text{d}N_{i}}{\text{d}E}}_{\text{Particle physics}},
\end{equation}
\noindent
where $ \langle\sigma v\rangle $ is the thermally-averaged annihilation cross section, $ m_{\chi} $ is the mass of the branon DM particle and $ \text{l.o.s.} $ stands for line-of-sight. The left panel of Fig.~\ref{fig:Branching_ratios_and_spectra} shows the branon branching ratios $ \text{Br}_{i} $ as a function of $ m_{\chi} $. The differential photon yield per branon annihilation $ \text{d}N/\text{d}E $ is depicted for a set of branon DM masses (from light to dark: 0.1, 0.2, 0.5, 1, 2, 5, 10, 20, 50 and $ \unit[100]{TeV} $) in the right panel of Fig.~\ref{fig:Branching_ratios_and_spectra}.

In our analysis, we are using the \textit{J}-factor and its statistical uncertainty for Segue~1 from~\cite{2015ApJ...801...74G} with the value of $ \mathrm{log}_{10} \left( J \left( \theta \right) [\unit{GeV^{2} cm^{-5}}] \right) = 19.02_{-0.35}^{+0.32} $. The  differential gamma-ray yields per annihilation $ \text{d}N_{i}/\text{d}E $ (see Eq. \ref{eq:Branon_Flux}) are taken from the PPPC 4 DM ID distribution~\cite{2011JCAP...03..051C}.

\section{Likelihood analysis method}
The data reduction of the Segue~1 observation have been published by the MAGIC Collaboration~\cite{2014JCAP...02..008A,2016JCAP...02..039M}. We re-analyse those high-level data in the context of brane-world extra-dimensional theories using the likelihood analysis described in Aleksi\'{c}, Rico and Martinez~\cite{2012JCAP...10..032A}. The binned ($ N_{\text{bins}} = 30 $) likelihood function of the dataset $ \bm{\mathcal{D}} $ with nuisance parameters $ \bm{\nu} $ reads as:
\begin{equation}
    \label{eq:Binned_lkl}
    \begin{split}
    \mathcal{L}_{\text{bin}} \left( \langle \sigma v \rangle ; \bm{\nu} \mid \bm{\mathcal{D}} \right) &= \mathcal{L}_{\text{bin}}( \langle \sigma v \rangle; \{ b_{i} \}_{i=1,\ldots,N_{\text{bins}}}, \tau \mid \{ N_{\text{ON},i}, N_{\text{OFF},i} \}_{i=1,\ldots,N_{\text{bins}}}) \\
    &= \prod_{i=1}^{N_{\text{bins}}} \Big[ \mathcal{P} (s_{i}(\langle \sigma v \rangle) + b_{i} \mid N_{\text{ON},i}) \cdot \mathcal{P} (\tau b_{i} \mid N_{\text{OFF},i}) \Big] \times \mathcal{T} \left( \tau \mid \tau_{\text{o}}, \sigma_{\tau} \right)
    \end{split}
\end{equation}
\noindent
where $ \mathcal{P} (x | N) $ is the Poisson distribution of mean $x $ and measured value $ N $ and $ s_{i}(\langle \sigma v \rangle) $ and $ b_{i} $ are the expected numbers of signal and background events in the $ i $-th energy bin. Besides $ b_{i} $, the normalization between background and signal regions $ \tau $, described by the likelihood function $ \mathcal{T} $, is also a nuisance parameter in the analysis~\cite{2022PDU....3500912A}. The total number of observed events in a given energy bin $ i $ in the signal (ON) and background (OFF) regions are $ N_{\text{ON},i} $, $ N_{\text{OFF},i} $, respectively.

The Segue~1 observational campaign $ \bm{\mathcal{D}}_{\mathrm{Segue1}} $ results in $ N = 8 $ distinct datasets with an individual set of instrument response functions (IRFs). The joint likelihood function
\begin{equation}
    \mathcal{L}\left( \langle \sigma v \rangle ; J, \bm{\nu} \mid \bm{\mathcal{D}}_{\text{Segue1}} \right) = \prod_{k=1}^{N} \Big[ \mathcal{L}_{\text{bin},k} \left( \langle \sigma v \rangle, \bm{\nu_{k}} \mid \bm{\mathcal{D}_{k}} \right) \Big] \times \mathcal{J} \left( J \mid J_{\text{o}}, \sigma_{\log_{10}J} \right)
\end{equation}
\noindent
is the product of the likelihood function of each dataset. We treat the \textit{J}-factor as a nuisance parameter and include the likelihood $ \mathcal{J} $ for the \textit{J}-factor~\cite{2015PhRvL.115w1301A}. $ \bm{\nu_{k}} $ represents the set of nuisance parameters different from the \textit{J}-factor affecting the analysis of the $ k $-th dataset $ \bm{\mathcal{D}_{k}} $.

The same analysis technique, implemented in the open-source DM analysis software packages~\cite{2021arXiv211201818M} \texttt{gLike}~\cite{Rico:gLike} and \texttt{LklCom}~\cite{Miener:LklCom}, has been used by the MAGIC Collaboration~\cite{2022PDU....3500912A,2018JCAP...03..009A} and other gamma-ray observatories in several DM searches~\cite{Oakes:2019,Armand:2021}.

\section{Results}

\begin{figure}[h]
    \centering
    \begin{subfigure}{.49\textwidth}
        \includegraphics[width=\textwidth]{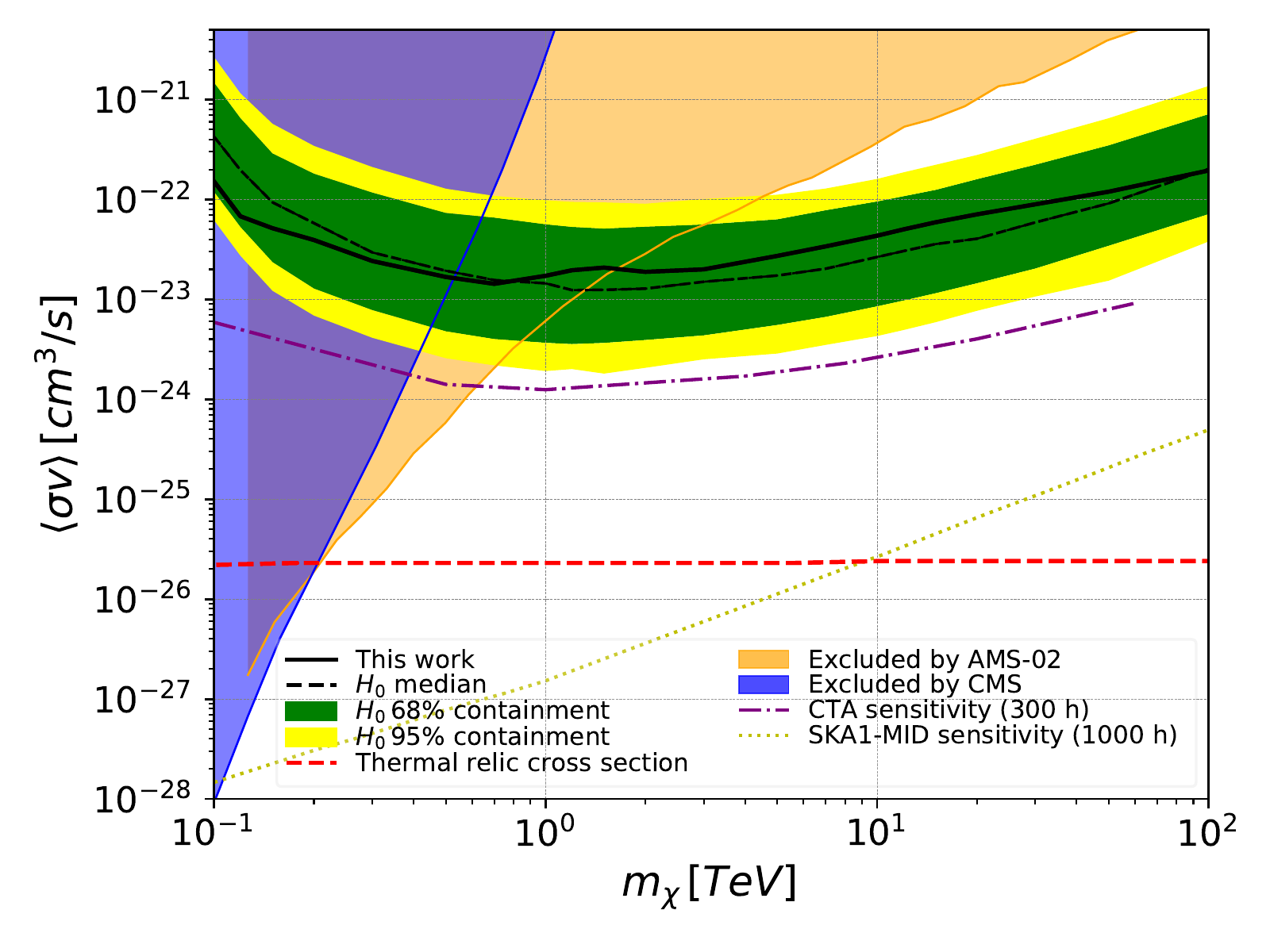}
    \end{subfigure}
    \begin{subfigure}{.49\textwidth}
        \includegraphics[width=\textwidth]{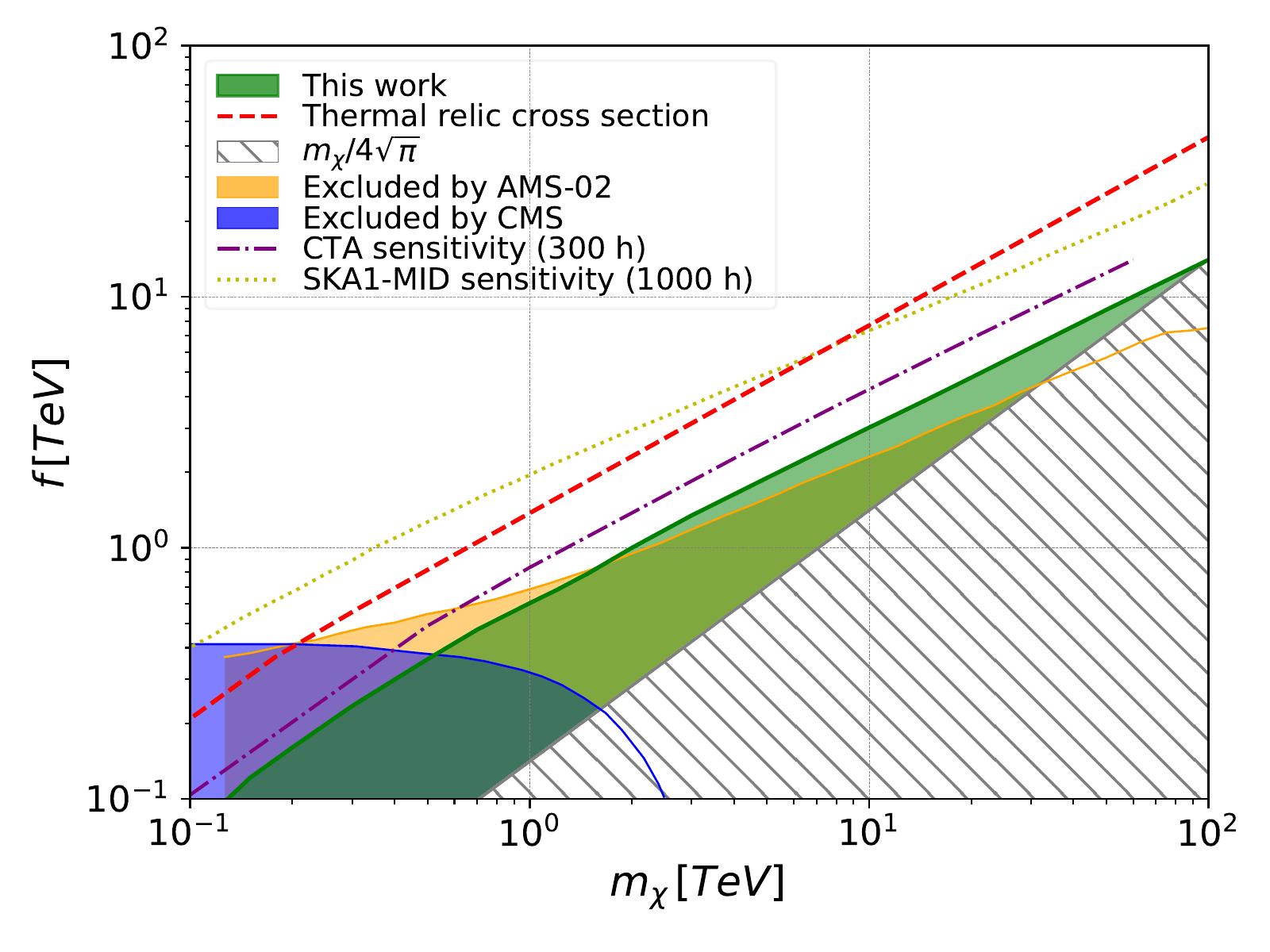}
    \end{subfigure}
\caption{Taken from JCAP05(2022)005. See text for more details.}
\label{fig:Limits}
\end{figure}

As reported in~\cite{2022JCAP...05..05}, we present the observational $ 95 \% $ confidence level (CL) upper limits on the thermally-averaged cross-section $ \langle \sigma v \rangle $ (see left panel of Fig.~\ref{fig:Limits}) and on the brane tension $ f $ (see right panel of Fig. \ref{fig:Limits}) for branon DM annihilation obtained with almost 160 hours of Segue~1 observation with the MAGIC telescopes. We perform a model dependent search for branon DM particles of masses between $ \unit[100]{GeV} $ and $ \unit[100]{TeV} $. Previous branon DM limits from CMS~\cite{2016physletb.755.102} (blue) and AMS-02~\cite{2019physletb.790.345} (orange) as well as the prospects of the future CTA~\cite{2020JCAP...10..041A} (purple) and SKA~\cite{2020PDU....2700448C} (yellow) are depicted in Fig.~\ref{fig:Limits}.

As expected from the no significant gamma-ray excess in the Segue~1 data~\cite{2014JCAP...02..008A}, our constraints for branon DM annihilation are located within the $ 68 \% $ containment band, which is consistent with the no-detection scenario. We obtain our strongest limit $ \langle \sigma v \rangle \simeq \unit[1.4 \times 10^{-23}]{cm^{3}s^{-1}} $ for a $ \sim \unit[0.7]{TeV} $ branon DM particle mass. 

\section{Conclusion and outlook}

This work leads to the most constraining branon DM limits in the multi-TeV mass range, superseding previous constraints by CMS~\cite{2016physletb.755.102} and AMS-02~\cite{2019physletb.790.345} above $ \sim \unit[1]{TeV}$. We can achieve even more stringent exclusion limits by adding further dSph observations of the MAGIC telescopes~\cite{2022PDU....3500912A} or other gamma-ray telescopes~\cite{Oakes:2019,Armand:2021} to this analysis.

\section*{Acknowledgments}
The authors thank the MAGIC Collaboration for providing private data.\newline
TM acknowledges support from PID2019-104114RB-C32. DN and TM acknowledge support from the former {\em Spanish Ministry of Economy, Industry, and Competitiveness / European Regional Development Fund} grant FPA2015-73913-JIN.\newline
VG’s contribution to this work has been supported by Juan de la Cierva-Incorporaci\'on IJC2019-040315-I grant, and by the PGC2018-095161-B-I00 and CEX2020-001007-S projects, both funded by MCIN/AEI/10.13039/501100011033 and by "ERDF A way of making Europe". VG thanks J.A.R. Cembranos for useful discussions. \newline
DK is supported by the European Union's Horizon 2020 research and innovation programme under the Marie Sk\l{}odowska-Curie grant agreement No. 754510. DK and JR acknowledge the support from MCIN/AEI/ 10.13039/501100011033 under grants PID2019-107847RB-C41 and SEV-2016-0588 ("Centro de Excelencia Severo Ochoa"), and from the CERCA institution of the Generalitat de Catalunya.

\end{document}